\documentclass[twocolumn,prb,showpacs,preprintnumbers,amsmath,amssymb]{revtex4}

\usepackage{graphicx}
\usepackage{dcolumn}
\usepackage{bm}

\begin{document}

\preprint{APS/123-QED}

\title{Multiorbital analysis of the effects of uniaxial and hydrostatic pressure on $T_c$ in the single-layered cuprate superconductors}

\author{Hirofumi Sakakibara$^{1}$}
\author{Katsuhiro Suzuki$^{1,6}$}
\author{Hidetomo Usui$^{2}$}
\author{Kazuhiko Kuroki$^{1,6}$}
\author{Ryotaro Arita$^{3,6,7}$}
\author{Douglas J. Scalapino$^{5}$}
\author{Hideo Aoki$^{4,6}$}

\affiliation{$\rm ^1$Department of Engineering Science, The University of Electro-Communications, Chofu, Tokyo 182-8585, Japan}
\affiliation{$\rm ^2$Department of Applied Physics and Chemistry, The University of Electro-Communications, Chofu, Tokyo 182-8585, Japan}
\affiliation{$\rm ^3$Department of Applied Physics, The University of Tokyo, Hongo, Tokyo 113-8656, Japan}
\affiliation{$\rm ^4$Department of Physics, The University of Tokyo, Hongo, Tokyo 113-0033, Japan}
\affiliation{$^5$Physics Department, University of California, Santa Barbara, California 93106-9530, USA}
\affiliation{$^6$ JST, TRIP, Sanbancho, Chiyoda, Tokyo 102-0075, Japan}
\affiliation{$^7$ JST, PRESTO, Kawaguchi, Saitama 332-0012, Japan}

\date{\today}

\begin{abstract}
The origin of uniaxial and hydrostatic pressure effects on $T_c$ 
in the single-layered cuprate superconductors is theoretically explored.  
A two-orbital model, derived from first principles and analyzed 
with the fluctuation exchange approximation gives 
axial-dependent pressure coefficients, 
$\partial T_c/\partial P_a>0$, $\partial T_c/\partial P_c<0$, 
with a hydrostatic response $\partial T_c/\partial P>0$  
for both La214  and Hg1201 cuprates, 
in qualitative agreement with experiments.  Physically, 
this is shown to come from a unified 
picture in which higher $T_c$ is achieved with an ``orbital 
distillation'', namely, the less the $d_{x^2-y^2}$ main band is 
hybridized with the $d_{z^2}$ and $4s$ orbitals higher the $T_c$.  
Some implications for obtaining higher $T_c$ materials are discussed.
\end{abstract}

\maketitle

\section{INTRODUCTION}

In the physics of high-$T_c$ cuprates, optimizing their $T_c$ 
remains a fundamental yet still open problem.   Empirically, important 
parameters that control $T_c$ have been identified for the cuprates, i.e., 
chemical composition, structural parameters, the number of layers, etc, 
besides the doping concentration.  
For the structural parameters specifically, 
several key quantities have been suggested: 
the bond length 
between copper and in-plane oxygen($l$, defined
 in Fig.\ref{fig1})\cite{Jorgensen,Bianconi}, and 
the Cu-apical oxygen distance ($h_{\rm O}$)
\cite{Maekawa,Andersen,Feiner,Pavarini,Kotliar,Weber,Takimoto,PRL,PRB}. 

Now, the pressure effect is exceptionally valuable as an 
{\it in situ} way to probe the structure-dependence of $T_c$. 
Regarding this, two general observations have been made 
for the cuprates: 
(i) $T_c$ tends to be enhanced under hydrostatic pressure, while 
(ii) uniaxial pressures produce anisotropic responses of $T_c$.  
More precisely, (i) $T_c$ has been shown to monotonically 
increase for pressure $< 30$ GPa \cite{Klehe,Gao}.  
As for (ii),  an $a$-axis 
compression generally raises $T_c$ 
($\partial T_c/\partial P_a>0$), 
while $c$-axis compression has an opposite 
effect ($\partial T_c/\partial P_c<0$)\cite{Hardy,Gugenberger,Meingast}.  Moreover, 
the magnitude of the pressure coefficient 
becomes smaller for materials 
having higher $T_c$'s, as summarized in 
Fig.3 of ref.\cite{Hardy}.   
The purpose of the present study is to theoretically reveal the 
origin of these general trends, focusing on the single-layered cuprates 
for clarity, and to shed light on a possibility of 
further optimizing $T_c$.

Conventionally, the theoretical model primarily used for the cuprates 
is a one-band Hubbard model based on the
$d_{x^2-y^2}$ orbital (or sometimes 
Cu-$3d_{x^2-y^2}+$O-$2p_{\sigma}$ orbital).  
Recently, we have shown\cite{PRL,PRB} 
that the $d_{z^2}$ orbital component 
strongly mixes into the states on the Fermi surface
in the relatively low-$T_c$ cuprates such as 
La$_2$CuO$_4$ (La214)\cite{Shiraishi,Eto,Freeman}, where the hybridization 
works destructively against $d$-wave superconductivity. 
While there have been some theoretical studies in the literature 
focusing on the role of 
the $d_{z^2}$ orbital\cite{Kotliar,Weber,Millis,Fulde,Honerkamp,Mori}, 
Refs.\cite{PRL,PRB} conclude that 
larger the level offset $\Delta E$ between the $d_{x^2-y^2}$ and $d_{z^2}$ 
Wannier orbitals, higher the $T_c$, where $\Delta E$ is governed by the
apical-oxygen height and the inter-layer distance.  
One might then presume that the effects of uniaxial 
pressures can simply be captured in terms of the pressure-dependence 
of $\Delta E$ affected by the crystal field.   
However, we reveal in the present work that 
the physics is not so simple.  
We find that, while the variation of $T_c$ under pressure 
is indeed affected by $\Delta E$, especially in the relatively low-$T_c$ 
cuprates, the large $\Delta E$ values in higher-$T_c$ cuprates such as 
HgBa$_2$CuO$_4$ (Hg1201) make their relevance to the $T_c$ variation smaller.   
We shall show that 
we have to turn our attention rather to 
the Cu$4s$ level, 
which is raised with pressure, 
resulting in a less rounded (i.e., better nested) Fermi surface. 
This, along with the increase in the band width, is shown to 
cause a higher $T_c$ under pressure.   
These results can be unified into a picture in which higher $T_c$ can be 
achieved by the ``distillation'' of the main (i.e., $d_{x^2-y^2}$) band, 
namely, the smaller the hybridization of other orbital components the 
better.

\section{FORMULATION}

\subsection{Construction of the two-orbital model}

Our theoretical procedure is as follows. 
We first determine the lattice structure under uniaxial and 
hydrostatic pressures from a 
first-principles band calculation with the Wien2k package\cite{wien2k}. 
From the band structure, we construct the maximally-localized 
Wannier orbitals\cite{MaxLoc,w2w} to obtain the hopping integrals 
for a two-orbital tight-binding model that takes into account both 
the $d_{x^2-y^2}$ and the $d_{z^2}$ Wannier orbitals explicitly\cite{PRL}.

\subsection{Many body analysis}

In this two-orbital model, we consider the onsite intra- 
and inter-orbital electron-electron repulsive interactions,
which are given, in the standard notation, as 
\begin{eqnarray}
H &=& \sum_i\sum_\mu\sum_{\sigma}\varepsilon_\mu n_{i\mu\sigma}
 + 
\sum_{ij}\sum_{\mu\nu}\sum_{\sigma}t_{ij}^{\mu\nu}
c_{i\mu\sigma}^{\dagger} c_{j\nu\sigma}\nonumber\\
&+&\sum_i\left( U\sum_\mu n_{i\mu\uparrow} n_{i\mu\downarrow}
+U'\sum_{\mu > \nu}\sum_{\sigma,\sigma'} n_{i\mu\sigma} n_{i\mu\sigma'}
\right.\nonumber\\
&-&\frac{J}{2}\sum_{\mu\neq\nu}\sum_{\sigma,\sigma'} c_{i\mu \sigma}^\dagger c_{i\mu \sigma'} 
c_{i\nu \sigma'}^\dagger c_{i\nu \sigma}\nonumber\\
&+&\left. J'\sum_{\mu\neq\nu} c_{i\mu\uparrow}^\dagger c_{i\mu\downarrow}^\dagger
c_{i\nu\downarrow}c_{i\nu\uparrow}
\right), 
\end{eqnarray}

\noindent where $i,j$ denote the sites while $\mu,\nu$ the two orbitals, 
the electron-electron interactions comprise 
the intraorbital repulsion $U$, interorbital repulsion 
$U'$, and the Hund's coupling $J$(= pair-hopping interaction $J'$ ).  
Here we take $U=3.0$ eV, $U'=2.4$ eV, and 
$J$=0.3 eV\cite{comment3}.  
These values conform to widely accepted, first-principles estimations 
 for the cuprates that $U$ is 7-10$t$ 
(with $t \simeq$ 0.45 eV), while $J, J' \simeq 0.1U$.   
Here we also observe the orbital SU(2) requirement, $U'=U-2J$.  

To study the superconductivity in this multi-orbital 
Hubbard model, we apply the fluctuation exchange approximation(FLEX)\cite{Bickers,Dahm,Kontani}.
In FLEX, we start with Dyson equation to obtain 
the renormalized Green's function, which is, in the multi-orbital case, 
a matrix in the orbital 
representation as $G_{l_1l_2}$, where $l_1$  and $l_2$ are orbital indices.
The bubble and ladder diagrams constructed from 
the renormalized Green's function are then summed to obtain the 
spin and charge susceptibilities, 
\begin{equation}
\hat{\chi}_s(q)=\frac{\hat{\chi}^0(q)}{1-\hat{S}\hat{\chi}^0(q)} ,
\end{equation}
\begin{equation}
\hat{\chi}_c(q)=\frac{\hat{\chi}^0(q)}{1+\hat{C}\hat{\chi}^0(q)} ,
\end{equation}
\noindent
where $q \equiv (\bm{q},i\omega_n)$ with 
wave vector $\bm{q}$ and with Matsubara frequency $i\omega_n \equiv (2n+1)\pi k_{\rm B} T$, 
 and the irreducible susceptibility is 
\begin{equation} 
\chi^0_{l_1,l_2,l_3,l_4}(q) =\sum_q G_{l_1l_3}(k+q)G_{l_4l_2}(k),
\end{equation}
with the interaction matrices 
\begin{equation}
S_{l_1l_2,l_3l_4}
=\left\{\begin{array}{cc}
U, &\;\; l_1=l_2=l_3=l_4\\ 
U',&\;\; l_1=l_3\neq l_2=l_4\\
J,&\;\; l_1=l_2\neq l_3=l_4\\
J',&\;\; l_1=l_4\neq l_2=l_3 ,
\end{array}  \right.
\end{equation}

\begin{equation}
C_{l_1l_2,l_3l_4}
=\left\{\begin{array}{cc}
 U &\;\; l_1=l_2=l_3=l_4\\ 
-U'+J & \;\; l_1=l_3\neq l_2=l_4\\
2U'-J,&\;\; l_1=l_2\neq l_3=l_4\\
J'& \;\; l_1=l_4\neq l_2=l_3 .
\end{array}  \right.
\end{equation}
With these susceptibilities, the fluctuation-mediated 
effective interactions are obtained, which are used to calculate the 
self-energy. Then the renormalized 
Green's functions are determined self-consistently from Dyson
equation. 
The Green's functions and the susceptibilities are 
used to obtain the spin-singlet pairing interaction in the form
\begin{equation}
\hat{V}^s(q)=\frac{3}{2}\hat{S}\hat{\chi}_s(q)\hat{S}
-\frac{1}{2}\hat{C}\hat{\chi}_c(q)\hat{C}+\frac{1}{2}(\hat{S}+\hat{C}),
\end{equation}
and this is used in the linearized Eliashberg equation,  
\begin{eqnarray}
\lambda \Delta_{ll'}(k)&=&-\frac{T}{N}\sum_q
\sum_{l_1l_2l_3l_4}V_{l l_1 l_2 l'}(q)\nonumber\\
&\times& G_{l_1l_3}(k-q)\Delta_{l_3l_4}(k-q)
G_{l_2l_4}(q-k).
\end{eqnarray}
The superconducting transition temperature, $T_c$, 
corresponds to the temperature at which the maximum eigenvalue $\lambda$ 
of the Eliashberg equation reaches unity, 
so that $\lambda$ at a fixed temperature can be used as a measure
for $T_c$.  
$T_c$ of the Hg cuprate is experimentally 
about three times higher than in La cuprate\cite{Eisaki}, 
so we calculate $\lambda$ by putting $T=0.01$ eV for La and 
$T=0.03$ eV for Hg for a clearer comparison. 
As we shall see, 
the eigenvalues discussed in the present study are away from unity 
(i.e., the temperature is higher than $T_c$)
due to the limitation in the 
number of Matsubara frequencies and the $k$-point meshes.
Therefore, for the La cuprate in particular, we restrict ourselves to 
qualitative argument for the 
$T_c$ variation under pressure.
For the Hg cuprate, on the other hand, 
we can go down to lower temperatures $(T\sim 0.01)$ where 
the eigenvalue approaches unity, 
and we have checked that the conclusions 
drawn from the $T=0.03$ calculation hold also for $T\sim 0.01$.
Moreover, we estimate $dT_c/dP$ for Hg with the $T\sim 0.01$ 
results, as will be discussed in the final part of the paper.
We fix the total band filling (number of electrons/ site) at $n=2.85$, 
for which the filling of the main band amounts to 0.85 (15 \% hole doping).
We take a $32\times 32\times 4$ $k$-point meshes for the three-dimensional lattice with $1024$ Matsubara frequencies.

\begin{table}[!t]
\caption{Structural and electronic parameters 
obtained from the first-principles(th) and  
experiments(exp) in ref\cite{La-st,Hg-st}.}
\label{table1}
\begin{tabular}{c| c c c c c}
\hline
 &\hspace{0.3cm}La(exp)\hspace{0.3cm}&\hspace{0.3cm}La(th)\hspace{0.3cm}&\hspace{0.3cm}Hg(exp)\hspace{0.3cm}&\hspace{0.3cm}Hg(th)\hspace{0.3cm}\\  \hline 
$a_0$ [\AA]            & 3.78 & 3.76  & 3.88 & 3.84 \\
$c_0$ [\AA]            & 13.2 & 13.1  & 9.51 & 9.58 \\
$h_{\rm O}$ [\AA]    & 2.42 & 2.41  & 2.78 & 2.81 \\
$h_{\rm La,Ba}$ [\AA] & 1.85  & 1.81  & 1.92 & 1.88 \\
$V_0$[\AA$^3$] &189 & 184  & 143 & 141 \\
$\Delta E$ [eV] &  0.857  & 0.861  &  2.16   & 2.305 \\
$r_{x^2-y^2}$         & 0.363 & 0.357  &   0.419  & 0.411 \\
$W$ [eV]        & 4.14  &  4.23   & 4.06 &  4.19 \\
\hline
\end{tabular}
\end{table}

\begin{figure}[!b]
\includegraphics[width=8cm]{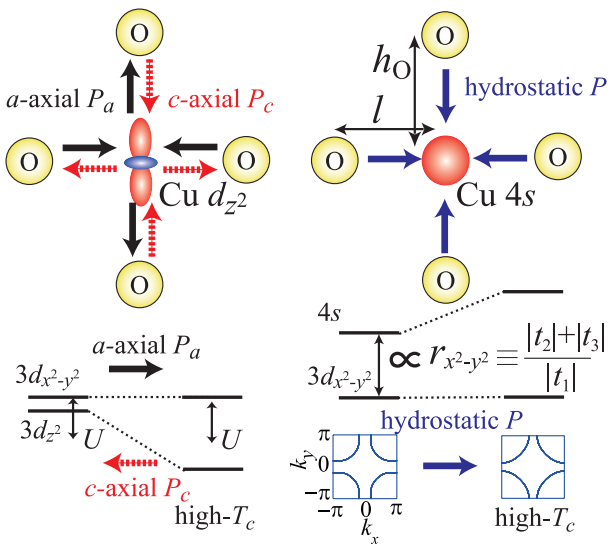}
\caption{Bottom left: schematic variation of the $d_{z^2}$ orbital level 
with respect to that for $d_{x^2-y^2}$ under uniaxial pressure (top left inset).  
Bottom right: the shift of Cu$4s$ level under hydrostatic pressure 
and its effect on the Fermi surface. 
}
\label{fig1}
\end{figure}

\section{CALCULATION RESULTS: UNIAXIAL PRESSURE}\label{Results}

\subsection{Crystal structure under pressure}

Let us begin with the case of uniaxial pressure. 
We first vary the lattice constants and calculate 
the total energy $E_{\rm tot}$. This is fit 
by the standard Burch-Marnaghan equation\cite{Birch} to 
determine the most stable structure 
with a unit cell volume $V=V_0$, the $a$-axis 
lattice constant $a=a_0$, and the $c$-axis $c=c_0$.  
For simplicity, we retain the tetragonal symmetry throughout, i.e., 
$b=a$ (so that the $a$-compression is actually biaxial).
We show in Table I the lattice parameters, $a_0$, $c_0$, 
$h_{\rm O}$, $h_{\rm La,Ba}$(La or Ba height 
measured from CuO$_{2}$ plane) and $V_{0}$, obtained for the La and Hg 
cuprates. The results are in good agreement with experimental values 
for the optimally doped compounds\cite{La-st,Hg-st}.  
We then relax the structure perpendicular to the compression direction, 
namely, we allow the lattice constant in that direction  to relax 
to obtain the value that gives the lowest energy.

\begin{figure}[!t]
\includegraphics[width=6cm]{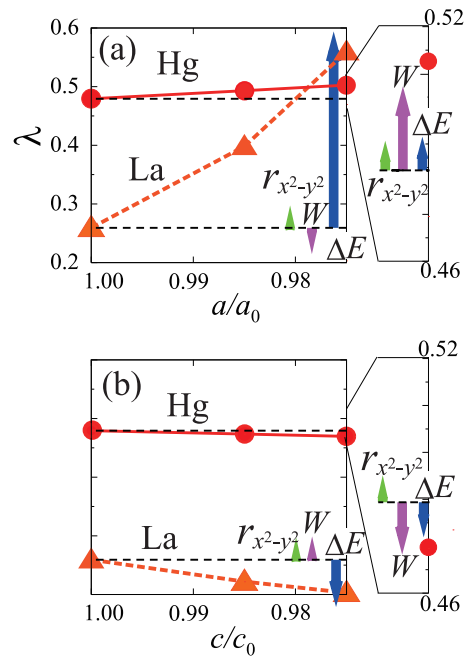}
\caption{
For uniaxial compressions the eigenvalue $\lambda$ of the Eliashberg equation is plotted against $a/a_0$ (a) or  $c/c_0$ (b).
Triangles (circles) indicate the result for the La (Hg) cuprates.
Arrows depict the contributions  (see text) 
to the $\lambda$ variation 
from $\Delta E$, $W$, and $r_{x^2-y^2}$, respectively,  at $a/a_0, c/c_0 = 0.975$.  
Lines are guide for the eye, with the dashed horizontal ones 
indicating the original values.
}
\label{fig2}
\end{figure}  

Figure \ref{fig2} plots the eigenvalue of the 
Eliashberg equation $\lambda$ against  
the lattice compression $a/a_0$  and $c/c_0$ for each compound.
The result shows that (i) in both compounds $\lambda$ increases as 
$a/a_0$ is reduced, while decreases as $c/c_0$ is reduced,
and (ii) the absolute value of the variations of $\lambda$ is larger in La 
than in Hg.  
These features are in qualitative agreement with the experimental 
observations  
summarized in Fig.3 of ref.\cite{Hardy}, which shows 
$\partial T_c/\partial P_a>0$ and 
$\partial T_c/\partial P_c<0$ for 
both materials, with larger $|\partial T_c/\partial P_i|$ in 
La than in Hg\cite{Hardy,Gugenberger}.
To be more precise, while the compressibility in the $a$ 
direction is nearly the same between the 
two materials, that in the $c$ direction is about three times larger in Hg 
than in La\cite{Takahashi,Balagurov} 
$(dc/dP_c|_{\rm Hg}\simeq 3 dc/dP_c|_{\rm La})$, 
but even if we take this into account, 
we find that 
$\partial \lambda/\partial P_c$ is still larger for La than for Hg 
in our calculation.

\subsection{Contribution from the $d_{z^2}$ orbital: $\Delta E$}

Now we want to pinpoint the origin of 
this $T_c$ variation against uniaxial pressures.  
In both materials, $\Delta E \equiv E_{dx^2-y^2} - E_{dz^2}$ 
increases as 
$a/a_0$ is reduced, while it decreases when $c/c_0$ is reduced.  
This is natural, since the $a$ $(c)$ reduction 
pushes the in-plane (out-of-plane) ligands toward 
Cu, resulting in a larger (smaller) crystal-field splitting and 
hence larger (smaller) $\Delta E$\cite{PRB}, as schematically depicted 
in Fig.\ref{fig1}.  
One might then expect that this alone is the 
origin of the $T_c$ variation, since $\Delta E$ and $T_c$ are positively correlated\cite{PRL}. To see if this is indeed the 
case, we have considered a case where we 
increase $\Delta E$ alone to its value at $a/a_0=0.975$ or $c/c_0=0.975$,
and obtain $\lambda$ with FLEX. The results are indicated in 
Fig.\ref{fig2} with arrows labeled as ``$\Delta E$''. In La, 
the resulting $\lambda$ is very close to those obtained for 
the actual compression, which implies that the main 
origin for $\lambda$ variation under uniaxial pressure comes from $\Delta E$.
By contrast, for Hg, the $\Delta E$ contribution is too small to 
account for the actual $\lambda$ variance (see the blowups in 
Fig.\ref{fig2}).

\subsection{Contribution from the $4s$ orbital: $r_{x^2-y^2}$ }

The reason for this is that in Hg, $\Delta E$
is $\simeq 2.5$ times larger than in La (table \ref{table1}), 
so that the effect of the $d_{z^2}$ 
orbital is tiny, while the contribution to the 
$T_c$ variation coming from other changes 
in the electronic structure become comparable with that from $\Delta E$.
In particular, we focus on 
the change in the energy difference $\Delta E_s$ between 
Cu$4s$ and Cu$d_{x^2-y^2}$ orbitals.
In fact, it has been shown that the Cu$4s$ orbital, 
which is implicitly included in the $d_{x^2-y^2}$ Wannier 
orbital in the present scheme,  
affects the second $(t_2)$ and third $(t_3)$ neighbor 
hoppings\cite{Pavarini,Andersen,PRL,PRB}. 
Note that the $4s$ orbital can be integrated 
out (implicitly included in the Wannier orbitals) 
prior to the many-body analysis, 
since the $4s$ orbital sits in energy well away from the Fermi level 
in contrast to the $d_{z^2}$ orbital (Fig.\ref{fig1})\cite{PRL,PRB}.
Smaller $\Delta E_s$ 
results in larger $r_{x^2-y^2} \equiv (|t_2|+|t_3|)/|t_1|$  
within the $d_{x^2-y^2}$ orbital sector, resulting in a 
more rounded Fermi surface, which degrades 
$d$-wave superconductivity\cite{Scalapino,PRL,Kent,Maier}.

\begin{figure}[!b]
\includegraphics[width=8cm]{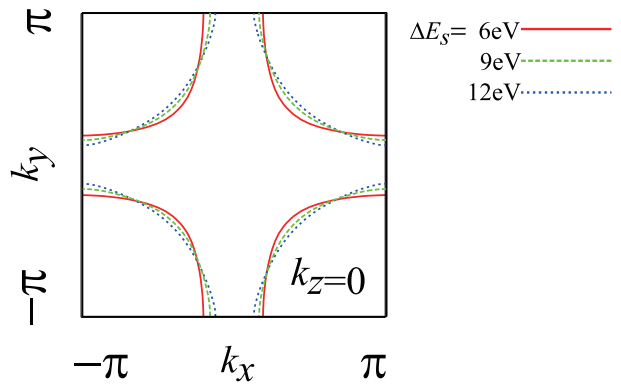}
\caption{
The Fermi surface of the three-orbital model of the Hg cuprate 
for values of 
$\Delta E_s$ hypothetical varied from 6eV (nearly original value) to 12eV.
}
\label{fig3}
\end{figure}

To show how the roundness varies with $\Delta E_s$, 
we consider a three-orbital model which explicitly 
includes  the Cu 4s Wannier orbitals for the Hg cuprate\cite{PRL,PRB},
and show in Fig.\ref{fig3} 
the Fermi surface for various values of 
$\Delta E_s= E_{{\rm Cu} 4s} - E_{{\rm Cu}3d_{x^2-y^2}}$.
We stress here that, while 
larger $\Delta E$ and larger $r_{x^2-y^2}$ (or smaller $\Delta E_s$) 
both give more 
rounded Fermi surface, their effects on $T_c$  are opposite.
Under pressure, $\Delta E_s$ is enhanced, which in turn 
reduces $r_{x^2-y^2}$.
In Fig.\ref{fig2}, we show the effect of hypothetically 
reducing $r_{x^2-y^2}$ down to its values at $a/a_0=0.975$ or $c/c_0=0.975$.
While the effect of $r_{x^2-y^2}$ is much smaller than that of $\Delta E$ in La,
the two effects are found to be comparable in Hg.

\subsection{Contribution from the band width: $W$}

In addition to $\Delta E$ and $r_{x^2-y^2}$, the band width $W$ 
(the energy difference between ${\bm k}=(0,0)$ and $(\pi,\pi)$)
 of the main band is also altered by pressure.  
In La the change in $\lambda$ due to the modification of 
$W$ is small compared to that 
arising from $\Delta E$, but in Hg the $W$ contribution 
is comparable with those from 
$\Delta E$ and $r_{x^2-y^2}$, which in fact 
provides a full understanding of the 
net $\lambda$ variation under uniaxial pressure.  
Namely, the $a$ $(c)$ reduction results in an increase (decrease) of the 
band width as expected, which enhances (suppresses) $T_c$. 
The increase of the band width results in a suppression of $U/W$, 
hence the electron 
correlation effect.  This reduces the pairing interaction, while the 
self-energy correction due to the spin fluctuations is reduced at 
the same time.  
The former has an effect of enhancing $T_c$, while the latter suppresses 
superconductivity.
In the case of Hg compound, the latter effect 
supersedes the former, resulting in an enhanced $T_c$.

It should be noted that 
the contribution from $r_{x^2-y^2}$, while relatively small 
for uniaxial compression,  
enhances $T_c$ for  {\it both} of the $a$- and $c$- axis compressions 
in marked contrast with the contributions from 
$\Delta E$ and $W$. This will become 
important in our analysis for hydrostatic pressures below.

\begin{figure*}[htbp]
\includegraphics[width=0.75\hsize]{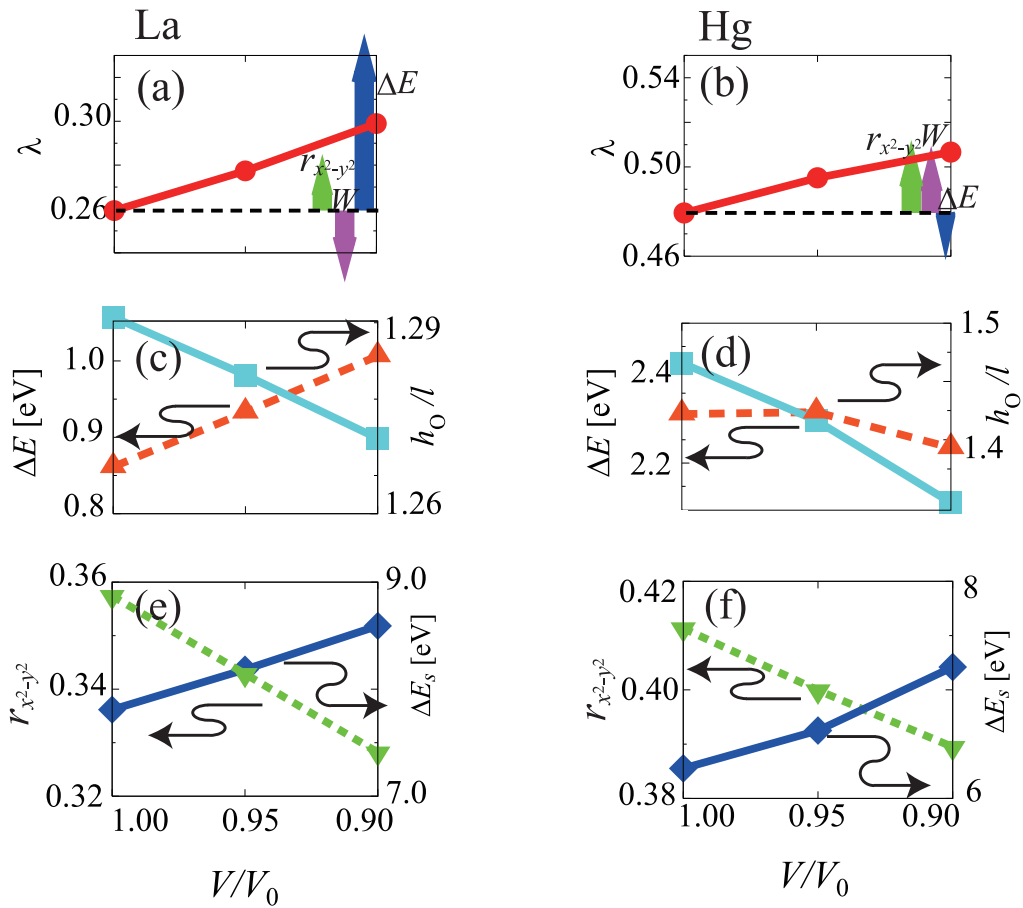}
\caption{
For hydrostatic pressure applied to La(Hg) cuprates in the left(right) column, 
(a),(b): The eigenvalue $\lambda$ of the Eliashberg equation plotted against the volume compression $V/V_0$.   
Arrows are as in Fig.\ref{fig2} for $V/V_0=0.90$.   
(c),(d): the value of $h_{\rm O}/l$(squares) and 
$\Delta E$(triangles).  
(e),(f): the value of $r_{x^2-y^2}$(triangles) and the 
$\Delta E_s 
\equiv E_{ {\rm Cu} 4s}
 - E_{{\rm Cu }
 3d_{x^2-y^2} }
$(diamonds).  
Lines are guide for the eye.}
\label{fig4}
\end{figure*}  

\section{CALCULATION RESULTS: HYDROSTATIC PRESSURE}

\subsection{La$_2$CuO$_4$}
Having identified the ingredients that determine the 
$T_c$ variation against uniaxial pressures, let us now move on to 
{\it hydrostatic} compression.  
Here we optimize the lattice structure at a fixed unit cell volume $V(<V_0)$
by varying Poisson's ratio, 
which we fit to the Burch-Marnaghan equation to obtain the most stable 
$E_{\rm tot}$.  Notably enough, for hydrostatic pressures 
$\lambda$  in Fig.\ref{fig4} increases with the volume compression 
in {\it both} materials.  This result 
qualitatively agrees with experimental results\cite{Klehe,Gao}.  
To understand its mechanism, we can, 
as done above for uniaxial pressures,  decompose 
the pressure effect on $\lambda$ into the contributions from $\Delta E$, $W$, and $r_{x^2-y^2}$ (arrows in Fig. \ref{fig4}(a)(b)).  
We can then realize that  the variation of $\Delta E$ 
against hydrostatic pressure is not as straightforward 
as in uniaxial pressures.  
Namely, we can look at $\Delta E$ along with 
the ``aspect ratio" 
$h_{\rm O}/l$
against the volume reduction in Fig.\ref{fig4}(c)(d), 
where $h_{\rm O}$ is the apical oxygen height and $l$ 
the in-plane Cu-O distance. 
Under hydrostatic pressure, $h_{\rm O}/l$ decreases 
in both materials 
because of the larger compressibility along the 
$c$ direction. One might then expect that this would reduce the crystal field 
splitting and hence $\Delta E$, but actually this is 
by no means always the case. 
In fact, $\Delta E$ increases with pressure for La, which is because the 
Cu-O distance decreases, 
resulting in a larger crystal-field effect. 
Thus the $T_c$ enhancement in La 
mainly comes from the increase of $\Delta E$. 

\subsection{HgBa$_2$CuO$_4$}

The above argument for La does not directly apply to Hg, since 
the original apical-oxygen height is larger, 
so that there is more room for the CuO 
octahedron to shrink along the $c$-axis 
than in La. 
Therefore, the $h_{\rm O}/l$ reduction is larger, 
resulting in a nearly constant $\Delta E$ against the 
decrease of $V/V_0$. This further makes $\Delta E$ irrelevant to 
the $T_c$ variation in Hg under hydrostatic pressures.  
As seen in Fig.\ref{fig4} with arrows, main 
contributions  to the $T_c$ enhancement 
come from $W$ and $r_{x^2-y^2}$, with similar magnitudes. 
As shown in Fig.\ref{fig4}(e)(f), the decrease of $r_{x^2-y^2}$ 
originally comes from an increase of the 
level offset $\Delta E_s$ introduced above.
The relatively large enhancement of the Cu$4s$ level under 
hydrostatic pressure can be understood from Fig.\ref{fig1} (right), 
where all the ligands approaching Cu push up 
the energy level of the extended and isotropic Cu$4s$ orbital 
to a larger extent than for 
the localized and anisotropic Cu$3d$ orbitals.  
Thus a message here is the {\it hydrostatic and uniaxial pressures exert 
significantly different effects}.   Specifically, the importance of $r_{x^2-y^2}$ becomes prominent in Hg 
in hydrostatic pressure because the $r_{x^2-y^2}$-contribution is  positive for both of $a$- and $c$-axis compressions, 
while $W$-contribution has opposite effects as shown in 
Fig.\ref{fig2}.


As for the band-width effect, we have found here 
that Hg exhibits an effect opposite to La for the present 
electron-electron interaction strength. To elaborate this, 
we have performed a FLEX calculation for various interaction strengths 
over $6<U/t<10$, and found that increasing the band width always 
results in an enhanced $\lambda$ in Hg within the considered 
compression range, while in La a similar effect 
is obtained only for $8<(U/t)$, with the effect reversing for 
smaller $U$.  We have further noticed that this 
``sign change'' in  the band-width effect against $U$ is peculiar to 
the systems having smaller $\Delta E$.  
At any rate, the band-width effect 
is much smaller than the effect of $\Delta E$ in La, so that 
the effect of pressure-dependence of $U$ 
does not affect the present conclusion.

\subsection{Order of magnitude of $dT_c/dP$}

Let us finally comment on the 
relation between the $\lambda$ variation for hydrostatic 
pressures and the $T_c$ enhancement in the 
actual pressure experiments. To see this, we 
have extended our calculation to lower temperatures 
for Hg, where $\lambda$ becomes closer to unity 
(i.e., $T$ approaches $T_c$). 
We find $\lambda\simeq 0.86$ at $T=0.01$ eV for $V=0.9V_0$, and the 
same value of $\lambda$ attained at $T=0.0088$ eV 
for $V=V_0$, so the temperature 
difference (a rough estimate of $\Delta T_c$) amounts to 
$\simeq 14$ K.   Since 
the compressibility is $\sim 0.01$ GPa$^{-1}$\cite{Balagurov}, 
this implies $dT_c/dP\sim 1$ K/GPa, 
which has the same order of magnitude 
found experimentally\cite{Hardy}.

\section{CONCLUDING REMARKS}

To summarize, we have identified the parameters that govern 
the $T_c$ variation of the single-layered cuprates under pressure. 
For lower-$T_c$ materials with small $\Delta E$ as exemplified 
by La$_2$CuO$_4$, $T_c$ is sensitive to 
$\Delta E$, which is identified to be the main contribution.  
For higher-$T_c$ materials with large 
$\Delta E$ as exemplified by HgBa$_2$CuO$_4$, $T_c$ is rather insensitive to $\Delta E$, and important 
contributions are revealed to come instead from the 
Fermi surface roundness governed by the Cu$4s$ orbital  
as well as the variation of the band width $W$. 
These effects coming from the electronic structure 
in the multi-orbital systems 
can be unified into a single picture in which the 
orbital distillation of the main band results in a higher $T_c$. 

The present study can also shed light on a 
materials-science avenue for optimizing $T_c$. The strategy for 
enhancing $T_c$, as conceived here, is: 
(1) keep the level offset  between the $d_{x^2-y^2}$ and 
$d_{z^2}$  orbitals large 
(ideally, larger than $U$ as shown in Fig.\ref{fig1} left),  
(2) expand the level offset between the Cu$4s$ 
 and the Cu$3d_{x^2-y^2}$ as much as possible ---  
this makes the Fermi surface more nested (fig.\ref{fig1} right), 
and 
(3) tune the band width $W$ to a moderate value.  
In this sense, it is important to keep the distance $h_{\rm O}$ between 
apical oxygen and Cu atom, 
and it is also important to decrease 
the in-plane Cu-O bond length $l$.  
In other words, the desired situation for 
optimizing $T_c$  should have 
an $a-b$ biaxial chemical  pressure which 
reduces the length $l$ from those in existing compounds, 
with the value of $h_{\rm O}$ kept high.  
This may be coupled to 
the possibility of the level offset $\Delta E_s$ controlled
 independently of $\Delta E$ by tuning length $l$\cite{PRL}.  

\section{ACKNOWLEDGMENTS}

The numerical calculations were performed at the Supercomputer Center, 
ISSP, University of Tokyo. This study has been supported by 
Grants-in-Aid for Scientific Research from JSPS
(Grants No. 23340095, RA;
No. 23009446, HS; No. 21008306, HU; and
No. 22340093, KK and HA). RA acknowledges financial support from 
JST-PRESTO.
DJS acknowledges support from the Center for 
Nanophase Material Science at Oak Ridge National Laboratory.

\end{document}